\documentclass[fontsize=11pt,twoside=semi,numbers=endperiod]{scrartcl}
\KOMAoptions{headinclude=true,footinclude=false,index=toc}
\usepackage[paperwidth=150mm,paperheight=179.35mm,width=110mm,totalheight=165.35mm,includehead,top=4mm,inner=20mm]{geometry} 

\usepackage{amsmath}
\usepackage{amssymb}

\usepackage{lmodern} 

\usepackage[ansinew]{inputenc}
\usepackage[T1]{fontenc}
\usepackage[english,ngerman]{babel}
\usepackage{textcomp} 
\usepackage{pifont} 
\usepackage[pdftex]{color,graphicx} %
\usepackage[rel]{overpic} 
\usepackage{bbm}
\usepackage{cite} 
\usepackage{ellipsis} 
\usepackage{microtype} 
\usepackage{simplewick} 
\usepackage{scrpage2}
\usepackage[%
	pdftex,
	colorlinks,citecolor=blue,filecolor=blue,linkcolor=blue,urlcolor=blue,%
	bookmarks=true,%
	bookmarksopen=false,%
	bookmarksnumbered=true,%
	pdfpagemode=UseNone,
    pdfstartview={XYZ null null 1.00},
	pdffitwindow=true,
    pdfview={XYZ null null null},
    pdfpagelayout=SinglePage,
    pdfdisplaydoctitle=true, 
    plainpages=false,
    pdfpagelabels=true,
	pdfkeywords={equivalence principle, Larmor's formula, free falling charge},
	pdftitle={Electrical charges in gravitational fields, and Einstein's equivalence principle},
    pdfauthor={Astrophysical Institute Neunhof}]
	{hyperref}

\setkomafont{disposition}{\rmfamily} 
\addtokomafont{section}{\large\bfseries} 
\addtokomafont{subsection}{\normalsize\bfseries} 
\addtokomafont{pagenumber}{\LARGE}
\addtokomafont{caption}{\small}

\deffootnote[1em]{1em}{1em}{\textsuperscript{\thefootnotemark}\ }

\newenvironment{ggitemize}{\vspace{0.2\baselineskip} \begin{addmargin}[2em]{0em}}{\vspace{-0.8\baselineskip} \end{addmargin}}
\newcommand{\ggitem}[2][$\ast $]{\makebox[0em][r]{#1\ \ }{#2}\newline }%

\newcommand{\ggie}{\mbox{i.\,e.}\ } 
 
\newcommand{\ggeg}{\mbox{e.\,g.}\ } 
 
\newcommand{\bz}[1]{\nolinebreak\hspace{0em}\nolinebreak{}#1\hspace{0em}}
\newcommand{\al}{\iflanguage{english}{``\nolinebreak\hspace{0em}}{\glqq\nolinebreak\hspace{0.1em}}\nolinebreak}
\newcommand{\ar}{\nolinebreak\iflanguage{english}{\hspace{0em}\nolinebreak{}''\ }{\hspace{-0.45em}\nolinebreak\grqq\thickspace}}
\newcommand{\arp}{\nolinebreak\iflanguage{english}{\hspace{0em}\nolinebreak{}''}{\grqq}}

\newcommand{\ggstackrel}[2][0]{\hspace{0.36em}\hspace{#1em}&=\hspace{-#1em}\hspace{-1.16em}\stackrel{#2}{\phantom{=}}}

\newcommand{\inte }{\int \limits }

\newcommand{\difq}[2]{\ensuremath{\frac{\mathrm{d}\hspace{0.1em}#1}{\mathrm{d}\hspace{0.1em}#2}}}

\DeclareMathOperator{\dif }{d}

\clearscrheadfoot 
\rohead[\vspace{-2mm}\par\pagemark ]{\vspace{-2mm}\par\pagemark}
\lehead[\vspace{-2mm}\par\pagemark ]{\vspace{-2mm}\par\pagemark}
\rehead{ } 
\lohead{ } 

\begin{document}
\selectlanguage{english}
\thispagestyle{empty} 
\raggedbottom  
\vspace*{3mm}
\centerline{\bfseries\Large Electrical charges in gravitational fields,}
\vspace*{.5mm}
\centerline{\bfseries\Large and Einstein's equivalence principle}
\vspace*{2mm}
\centerline{{\bfseries Gerold Gr\"{u}ndler}\,\footnote{\href{mailto:gerold.gruendler@astrophys-neunhof.de}{gerold.gruendler@astrophys-neunhof.de}\hfill\href{http://www.astrophys-neunhof.de/}{www.astrophys-neunhof.de}}}
\vspace*{1mm}
\centerline{\small Astrophysical Institute Neunhof, N\"{u}rnberg, Germany} 
\vspace*{3mm}
\noindent\parbox{\textwidth}{\small According to Larmor's formula, accelerated electric charges radiate electromagnetic waves. Hence charges should radiate, if they are in free fall in gravitational fields, and they should not radiate if they are supported at rest in gravitational fields. But according to Einstein's equivalence principle, charges in free fall should not radiate, while charges supported at rest in gravitational fields should radiate. In this article we point out indirect experimental evidence, indicating that the equivalence principle is correct, while the traditional interpretation of Larmor's formula must be amended.} 
\vspace*{1mm}

\section{Einstein's Equivalence Principle (EP)} 
Due to experimental evaluations and due to gedanken\bz{-}experiments, Galilei demonstrated\!\cite{Galilei:discorsi} that all bodies, independent of their materials and independent of their weights, would fall down to earth with equal acceleration, if air friction could be eliminated. 

In Newton's theory\!\cite{Newton:Principia}, Galilei's finding is a simple consequence of the proportionality
\begin{subequations}\label{knbgxmybg}\begin{align} 
m_{\text{inertial}}=\text{constant}\cdot m_{\text{gravitational}} 
\end{align}
between inertial and gravitational mass. The constant is independent of the materials and weights of masses, and can be chosen by definition as 
\begin{align} 
\text{constant}&\equiv 1\\ 
\Longrightarrow\quad m&\equiv m_{\text{inertial}}=m_{\text{gravitational}} \ . 
\end{align}\end{subequations}
As a consequence of \eqref{knbgxmybg}, if the force 
\begin{align} 
\boldsymbol{F}=m\boldsymbol{a} 
\end{align}
is acting on each body in a laboratory, then it is impossible to find out by any mechanical measurement inside the laboratory (without looking out of the windows) whether the laboratory is at rest within a homogeneous gravitational field, which is exerting the gravitational acceleration 
\begin{subequations}\begin{align} 
\boldsymbol{g}=+\boldsymbol{a}=\frac{\boldsymbol{F}}{m} \ ,\label{ksdjdjvjna} 
\end{align} 
or whether the laboratory is --- far\bz{-}off any measurable gravitational field --- being boosted by a rocket with acceleration 
\begin{align} 
\boldsymbol{a}_{\text{boost}}=-\boldsymbol{a}=-\frac{\boldsymbol{F}}{m}\ .\label{ksdjdjvjnb} 
\end{align}\end{subequations} 
Obviously, $\boldsymbol{g}=\eqref{ksdjdjvjna}$ can be constant (within measurement accuracy) only in sufficiently small laboratories, while in larger laboratories tidal effects and other inhomogeneities will be observable. 

Einstein postulated\!\cite{Einstein:GravLicht} the perfect equivalence of accelerated laboratories and sufficiently small inertial laboratories in gravitational fields with regard not only to mechanical phenomena, but to \emph{all} physical phenomena --- and hence to \emph{all} laws of nature. 
\begin{align} 
\mbox{\parbox{.86\linewidth}{Einstein's Equivalence Principle (EP)\,:\newline{}All laws of nature are identical in an inertial reference system in a homogeneous gravitational field with gravitative acceleration \raisebox{0mm}[0mm][0mm]{$\boldsymbol{g}$}, and in a reference system which is accelerated by \raisebox{0mm}[0mm][0mm]{$\boldsymbol{a}_{\text{boost}}=-\boldsymbol{g}$} in a region of space which is free of measurable gravitation.}}\label{eq:EP}  
\end{align}
The EP is one of the main pillars, onto which Einstein founded his General Relativity Theory (GRT). 

\section{Accelerated electrical charges} 
While the EP is correct beyond doubt for mechanical phenomena (as is well\bz{-}known since Newton's days), there is an ongoing dispute since decades, whether the EP is correct with regard to accelerated electrical charges, which are (or are not?) emitting electromagnetic radiation. Based on Maxwell's classical theory of electromagnetism\!\cite{Maxwell:treatise}, Larmor\!\cite{Larmor:radform} computed the electromagnetic power radiated by a particle with charge $q$ and velocity $\boldsymbol{v}$, which is accelerated by $\boldsymbol{\dot{v}}$ (note that SI units\!\cite{SI:units} are used throughout this article):   
\begin{subequations}\label{ksahghsdg}\begin{align}
P=\frac{2q^2\,\boldsymbol{\dot{v}\cdot\dot{v}}}{3c^3(4\pi\epsilon _0)}\quad\text{if }v\ll c\label{ksahghsdga}
\end{align} 
With $m$ being the particle's rest mass, and $\dif\! p/\dif\!\tau $ being the derivative of it's four\bz{-}momentum $(p ^\mu )\equiv (\gamma mc,\gamma m\boldsymbol{v})$ with respect to it's proper time $\tau $, the relativistic generalization of \eqref{ksahghsdga} is 
\begin{align}
P=-\frac{2q^2}{3c^3(4\pi\epsilon _0)m^2}\,\difq{p_\nu}{\tau }\,\difq{p^\nu}{\tau }\ . \label{ksahghsdgb}
\end{align}\end{subequations}
Both equations \eqref{ksahghsdg} are (only) valid in inertial reference systems. The derivation of \eqref{ksahghsdg} is demonstrated in very detail in \cite{apin:se90115}. The qualitative and quantitative correctness of \eqref{ksahghsdg} has been confirmed by countless technical applications like radio antennas, X\bz{-}ray tubes, and synchrotrons. 

\section{The contradiction}\label{sec:contradict}
Consider two identical charges, one held at rest some meters above earth surface, and the other falling down from some meters height to earth surface. As the Coriolis acceleration and tidal accelerations (by which a reference system fixed to earth surface differs from a true inertial system) are much smaller than the gravitational acceleration due to the mass of the earth, according to Larmor's law \eqref{ksahghsdg} the falling charge, being accelerated with $\dot{v}\approx 9.81\,\text{m}/\text{s}^2$, should radiate much stronger than the charge at rest relative to earth surface. 

According to Einstein's equivalence principle \eqref{eq:EP}, however, the charge at rest in the earth's gravitational field ($g\approx 9.81\,\text{m}/\text{s}^2$) should radiate, because the setup is equivalent to a setup in which the same charge is boosted with $\dot{v}\approx -9.81\,\text{m}/\text{s}^2$ in a region of space free of significant gravitation, while the charge in free fall should not radiate at all, because the setup is equivalent to the same charge being not boosted in a region of space free of significant gravitation. 

Thus the EP \eqref{eq:EP} and Larmor's formula \eqref{ksahghsdg} seem to be incompatible. Apparently at least one of them must be wrong. Below we will argue, however, that the EP is completely correct, and that Larmor's formula actually is not wrong, but must be interpreted differently than done above. 

During the decades, theorists have worked out highly sophisticated constructions, some of them (allegedly) proving that and why free falling charges radiate, while charges supported at rest in gravitational fields do not radiate, others of them (allegedly) proving that and why just the opposite is true. See \cite[section\,4]{Groen:equivprinc} for a review of the theoretical achievements. 

In a situation where theorists can not find consensus for decades, usually the experimentalists should provide the decision. But in this case that's easier said than done. 

\section{No help from direct experimental observation}\label{sec:dirobs} 
If a cloud of $N$ elementary charges \mbox{$e=-1.6\cdot 10^{-19}\text{C}$} is accelerated by $g\approx 9.81\,\text{m}/\text{s}^2$, then according to Larmor's law the power 
\begin{align}
P\stackrel{\eqref{ksahghsdga}}{=}N\cdot\frac{2e^2g^2}{3c^3(4\pi\epsilon _0)}\approx N\cdot 5.5\cdot 10^{-52}\,\text{W} 
\end{align} 
will be radiated. If we could measure a radiated power $\geq 1\,\text{pW}$ (which is a quite ambitious objective), we would need to let a charge as large as $-e\cdot 10^{40}$ fall down to earth, or hold it at rest above earth surface, to achieve a measurable result, {\small AND} at the same time we would need to make sure that the charge really is subject only to gravitation, but not to the electrostatic force exerted by the charge of $+e\cdot 10^{40}$, which remained on earth when the opposite charge was prepared. Considering that the earth is consisting of about $10^{50}$ atoms, and that the attraction between a charge $-e$ and a charge $+e$ is at same distance about $10^{33}$ times larger than the gravitative attraction between two silicon atoms, that experiment is clearly impossible on earth. 

Alternatively we might try to observe somewhere in the universe charges falling in gravitational fields, which are much stronger than the field of the earth. But if we are lucky and observe such radiation, it will hardly be possible to prove that the radiating charge, lightyears away from the observer, really is in free fall, and is not accelerated by some external electromagnetic field.  

Hence it is impossible by today, and will probably (disputably) stay impossible forever, to decide by direct experimental observation whether charges radiate or not, if they respectively are in free fall or supported at rest in gravitational fields. 

\section{But there is some indirect experimental evidence} 
The indirect evidence announced here, results simply from the analysis of radiating charges in antennas, X\bz{-}ray tubes, and synchrotrons. 

First consider the synchrotron. In storage rings, electrically charged particles are subject to an inertial centrifugal acceleration, which is exactly balanced by the centripetal acceleration exerted by the Lorentz force. Hence the orbiting charges feel weightless, like cosmonauts orbiting around the earth in the International Space Station feel weightless. Described in an inertial reference system, however, the charges are accelerated and should radiate. But how can the orbiting charges know that emittance of synchrotron radiation is due, even though they do not feel that acceleration? 

The obvious answer is: A charged particle indeed does not feel any radial acceleration in a storage ring, but the electromagnetic field, emanating from the particle, does. The electromagnetic field is not an integral part of the charged particle. Instead it is something external to the particle. There is energy stored in the field. Hence the field is subject to gravity and inertial forces. But it is not subject to the Lorentz force, which the magnetic field of the storage ring is exerting onto the charged particle. While the Lorentz force is precisely compensating the inertial centrifugal force acting onto the particle, the centrifugal force acting onto the field is not compensated by any centripetal force. Therefore the Lorentz force is producing a \emph{relative acceleration} between the stored elementary particle and it's electromagnetic field. 

In antennas and X\bz{-}ray tubes, just the opposite is true: Electrons are accelerated, and feel the acceleration, while their electromagnetic fields are not accelerated. 

In any case of radiating charges, there is a \emph{relative acceleration} between the charge and it's field. Nobody ever observed a charge radiating, while there was no relative acceleration between the charge and it's field. Hence we arrive at a consistent picture, which\vspace{-1\baselineskip} 
\begin{ggitemize}
\ggitem[(a)]{is in accord with all experimental observations of radiating charges, and}
\ggitem[(b)]{removes the apparent contradiction between Larmor's formula and the EP,}
\end{ggitemize}
if we assume that it is just the relative acceleration between a charge and it's field, which is causing the emission of radiation, and therefore clarify the correct interpretation of Larmor's formula by this 
\begin{align} 
\mbox{\parbox{.86\linewidth}{Application note for Larmor's formula \eqref{ksahghsdg}\,:\newline{}$\boldsymbol{\dot{v}}$ is not an arbitrary acceleration, but a relative acceleration between the charge and it's electromagnetic field, \ggie  an acceleration due to any force but not gravity.}}\label{eq:interpret}  
\end{align}
Indeed, if the derivation\!\cite{apin:se90115} of Larmor's formula is carefully reviewed, it becomes obvious that $\boldsymbol{\dot{v}}$ is always understood as an acceleration due to an external force, which does affect only the charged particle, but not the electromagnetic field emanating from it. Therefore $\boldsymbol{\dot{v}}$ must not be confused with gravitational acceleration. 

Stress is just another name for relative acceleration. The conviction, that stress between a charged particle and it's field is the true cause of the radiation, is not new. This approach has been worked out in particular by Soker and Harpaz\!\cite{Soker:radchargeerst,Soker:radcharge}. What we have done here, is essentially to point out experimental evidence (synchrotron radiation of charged particles in storage rings, as compared to radiating electrons in antennas and X\bz{-}ray tubes) in support of this point of view. 

\section{Charges supported at rest in gravitational fields}\label{sec:chargesupport}
This case is more complex than expected, because the intricate general\bz{-}relativistic effect of event\bz{-}horizons comes into play. First, we conclude from our previous discussion: A charge, supported at rest in a gravitational field, should radiate, because the support is holding the charge at it's place due to electromagnetic forces which exactly balance the gravitational force, while the field emanating from the charge is not supported, but subject to gravity. Hence there is stress between the particle and it's field, resulting into radiation. 

In apparent contradiction, Unnikrishnan and Gillies\!\cite{Unnikrishnan:radcharge} cite two convincing arguments, demonstrating that an observer supported at rest in the same gravitational field, will not observe radiation: First, there is no current and hence no magnetic field in this static setup, which would be an indispensable precondition for electromagnetic radiation. Second, energy conservation would be violated: The energy content of the gravitational field, and in particular the potential energy of the charge at constant height in the gravitational field, is constant while it is radiating. We could surround the radiating charge by antennas, let the received radiation do work, and thus have an inexhaustible source of energy, the perfect perpetuum mobile. 

But another observer, in free fall passing by the charge, could see the radiation, as is obvious if we apply the EP: If the support of the charge --- far off any measurable gravitational field --- is accelerated by a rocket drive, then in an inertial reference system (\ggie  in the system, in which the free falling observer is at rest) there is very well a current and hence a magnetic field. And the accelerated support very well is doing work, supplying the energy for acceleration of the charge and the radiation emitted by the charge. 

There is no true contradiction, however. In an important article, Rohrlich\!\cite{Rohrlich:equivalence} showed that the radiation, emitted by an accelerated charge, can be observed by a detector at rest in an inertial system (\ggie  accelerated relative to the emitting charge), while a detector, which is co\bz{-}moving with the accelerated charge, will only see an electrostatic field. Transformed back into the picture with the charge at rest in the gravitational field, this result says that the co\bz{-}moving detector, which as well is at rest in the gravitational field, will observe only an electrostatic field but no radiation, while another detector, which in accelerated movement (\ggeg  in free fall) is passing by the charge, will observe the radiation. According to GRT, there exists an event horizon --- often called Rindler\bz{-}horizon --- between the charge supported at rest in the gravitational field, and the observer at rest relative to the charge. The observer can not see the radiation, which is beyond his event\bz{-}horizon. The effect of event horizons has been evaluated in detail by Boulware\!\cite{Boulware:acccharge}. For a particular clear treatment, see the article by de\,Almeida and Saa\!\cite{Almeida:horizon}. 

Hence the perpetuum mobile, with it's antennas at rest relative to the charge, will not work. And antennas, passing by in free fall the charge at rest, wouldn't violate energy conservation, because they loose during their fall much more potential energy than they can gain due to absorption of radiation. 

Rohrlich's result has bee rejected by Soker and Harpaz\!\cite{Soker:radcharge}, who argue that emission and absorption of radiation are objective events, which can not simply disappear by whatever transformation of time\bz{-}space coordinates. Even if the co\bz{-}moving observer can not see the radiation, he can of course note, that a free falling observer is receiving energy. For example, the free falling observer could use the received energy to shine a light, and the co\bz{-}moving observer, at rest relative to the charge, could see the light. But according to observation of the co\bz{-}moving observer, no energy has been extracted from the electrostatic field of the charge. Isn't energy conservation violated? Clearly this question deserves a more detailed analysis. 

Let's build a  receiver device, consisting of an antenna and a battery, into which all the electromagnetic energy is stored which is received by the antenna. $M$ is the mass of the receiver, when the battery is empty. To carry this receiver from earth surface up to a tower of height $H_0$ above earth surface, the work 
\begin{subequations}\begin{align}
W_{\text{start}}=M\!\inte _0^{H_0}\!\dif\! h\,g(h) 
\end{align}
must be done. At height $H_0/2$ a charge is fixed to the tower. Now we let the receiver fall down. Being in free fall, the receiver will see the charge radiating, pick up the radiation energy $\Delta E$, and then crash onto earth surface. The energy $\Delta E$ can be extracted from the battery, and can be used to do some work. From point of view of the observer, who is at rest relative to the charge, no energy has been extracted from the electrostatic field of the charge. Therefore in his reference system, the energy is at the end 
\begin{align}
W_{\text{end}}=\Delta E +W_{\text{crash}}\ . 
\end{align}
Energy conservation requires 
\begin{align}
W_{\text{start}}=W_{\text{end}}\ . 
\end{align}\end{subequations}
Unfortunately this setup is not suited to check energy conservation, because the energy $W_{\text{crash}}$ is dissipated as uncontrolled heat into the environment. 

In a gedanken\bz{-}experiment, we can improve the setup and get rid of the crash\footnote{I thank Noam Soker (Technion, Haifa), who suggested in private communication this crash\bz{-}free setup.}: Where the receiver crashed to earth, we dig a hole right down to the center of the earth, and straight further to the antipode surface. 

In a first experimental run, we remove the charge, and let the receiver fall down from the tower (height $H_0$) into the hole. As air friction is negligible in this gedanken\bz{-}experiment, the receiver will swing through the earth, turn at height $H_0$ above antipode earth surface, and swing back to our surface up to height $H_0$.\footnote{Probably almost every physicist has computed this little exercise in the first months of his\bz{/}her studies.} The battery is empty, of course, because the charge has been removed. 

In a second experimental run, two identical charges are placed on earth surface diametrically opposed right and left nearby the hole entrance. Thus we make sure that the receiver will not get any transversal momentum, when it absorbs the radiation. Again we let the device fall down (with empty battery) from height $H_0$. When it has returned from it's round\bz{-}trip through the earth, the energy $E_{\text{battery}}>0$ can be extracted from the battery. As --- from point of view of the observer, who is at rest relative to the charges --- no energy has been extracted from the electrostatic field of the charges, energy conservation requires that the energy stored in the battery must come from the gravitational field. Hence we expect that the returning receiver will not reach height $H_0$ any more, but only height $H_1<H_0$, which is determined by 
\begin{align} 
E_{\text{battery}}=M\!\inte _{H_1}^{H_0}\!\dif\! h\, g(h)\ .\label{ishngsjdg} 
\end{align} 
A simple mechanical model, which doesn't require relativistic treatment and consideration of event horizons, may serve to make this result plausible: The charges are replaced by guns, which fire a bullet each into the receiver, when it passes by. Let $v_{\text{a}}$ be the velocity of the receiver just before it is hit by the bullets on it's way from the tower down into the hole. $v_{\text{b}}$ is it's velocity immediately after it absorbed the bullets. When returning after the round\bz{-}trip through the earth to the surface, the velocity of the receiver again is $v_{\text{b}}$. Then again two bullets are fired onto it, and $v_{\text{c}}$ is it's velocity immediately after it absorbed the bullets. 

With horizontal alignment of the guns, momentum conservation gives  
\begin{align} 
Mv_{\text{a}}=(M+2m_{\text{bullet}})v_{\text{b}}=(M+4m_{\text{bullet}})v_{\text{c}}\ .\label{mnsngnbs} 
\end{align} 
After the receiver was hit a second time by two bullets, it moves up to height $H_1$, converting all it's kinetic energy into potential energy: 
\begin{align} 
(M+&4m_{\text{bullet}})\!\inte _0^{H_1}\!\dif\! h\,g(h)\, =\notag\\ 
&=\frac{1}{2}\,(M+4m_{\text{bullet}})v_{\text{c}}^2\stackrel{\eqref{mnsngnbs}}{=}\frac{1}{2}\,\frac{M}{1+4m_{\text{bullet}}/M}\, v_{\text{a}}^2=\notag\\ 
&=M\!\inte _0^{H_0}\!\dif\! h\,g(h)\ -2m_{\text{bullet}}v_{\text{a}}^2+\mathcal{O}(m_{\text{bullet}}^2/M^2) 
\end{align} 
From this equation we read 
\begin{align} 
4m_{\text{bullet}}\!\inte _0^{H_1}\!\dif\! h\,g(h)&=M\!\inte _{H_1}^{H_0}\!\dif\! h\,g(h)\ -\notag\\ 
&-2m_{\text{bullet}}v_{\text{a}}^2+\mathcal{O}(m_{\text{bullet}}^2/M^2)\ .\label{ksmngvnsdngb}  
\end{align} 
The potential gravitative energy of the bullets in the mechanical model --- \ggie  the left side of this equation --- is resembling the electromagnetic energy $E_{\text{battery}}$ stored in the battery. The first term on the right side is the energy extracted from the gravitational field. In \eqref{ishngsjdg}, the gravitational field is the only source for the energy gain of the receiver. But in the mechanical model, there is a further source --- the second line in \eqref{ksmngvnsdngb} --- from which the receiver has harvested energy. This further source of energy are the guns (\ggie the chemical reactions in their cartridges), which add kinetic energy to the receiver: 
\begin{align} 
\Delta E_{\text{kin}}&=\frac{1}{2}\, (M+4m_{\text{bullet}})\, v_{\text{c}}^2-\frac{1}{2}\, Mv_{\text{a}}^2=\notag\\ 
\ggstackrel{\eqref{mnsngnbs}}\frac{1}{2}\,\Big(\frac{M}{(1+4m_{\text{bullet}}/M)}-M\Big)\, v_{\text{a}}^2=\notag\\ 
&=-2m_{\text{bullet}}v_{\text{a}}^2+\mathcal{O}(m_{\text{bullet}}^2/M^2) 
\end{align} 
In contrast, the radiation energy of the charges must come exclusively from the gravitational field. Besides this difference, \eqref{ksmngvnsdngb} may be considered a confirmation of \eqref{ishngsjdg}. 

Thus there is no conflict with energy conservation. Still the observer (at rest relative to the charges) may wonder what is going on: As he can --- in this gedanken\bz{-}experiment of classical physics --- perform measurements with arbitrary accuracy, he will note that the receiver is decelerated a little bit each time when it passes by the charges, see \eqref{mnsngnbs}. The observer will conclude that kinetic energy is transformed into electromagnetic energy, which he finds after the experiment in the battery. The kinetic energy of the receiver again results from it's potential energy in the earth's gravitational field. In total, gravitational energy is converted into electromagnetic energy, which is stored in the battery of the falling receiver. 

The amount of energy conversion is inversely proportional to the square of the distance between the charges and the receiver. If the charges are removed, no energy conversion happens at all. The presence of the charges obviously is indispensable to make the conversion of gravitational energy into electromagnetic energy happen, but still neither the charges nor their electrostatic fields seem to be affected by the process. The charges resemble the catalysts, which are applied in some chemical reactions. The catalysts don't seem to be involved in the chemical process, but still their presence is indispensable to bring about the reaction. 

At this point of our considerations, we see that we have not only reached, but actually already slightly violated the limits of the application range of classical physics. Firstly note, that \emph{all} energy radiated by charges at rest in gravitational fields must be absorbed by some accelerated receiver. If radiation would disappear un\bz{-}absorbed to infinity, then the radiated energy could not be replenished by gravitational energy due to deceleration of a receiver, and energy conservation would be violated. 

Note secondly: Energy radiated by the charges needs a finite time to propagate to the receivers. But extraction of energy from the gravitational field only starts, when the radiation is being absorbed by the receiver. In other words: The radiation reaction force, which is supplying the energy of the radiation, here is perplexingly working on the receiver, but not on the emitter of the radiation. Hence energy conservation is violated for the time interval between emission and absorption of radiation. 

Odd timing of energy transfer is a well\bz{-}known problem of the classical model of radiating charged particles, and the issue of radiation reaction never found a satisfying solution within classical physics. When Dirac considered the acceleration of an electron due to an electromagnetic force which is acting only for a short moment (\ggie  a pulse), he found\!\cite[equation (35)]{Dirac:classelectron} that the electron is already accelerated \emph{before} the pulse arrives at the particle's position. This is often called \al pre\bz{-}acceleration\ar  in the literature. Wrong timing of energy transfer is a general problem of the classical radiation model, not restricted to charges in gravitational fields. 

Wheeler and Feynman tried in their absorber theory\!\cite{Wheeler:absorber}, to cope with both problems. They assumed that \emph{all} radiation ever emitted in the universe is absorbed within finite time by an absorber. And they assumed that accelerated charges and absorbers are interacting both retarded and advanced, to overcome the timing problem. Their absorber theory suffers from severe flaws\!\cite{Gruendler:absorber}, however, and does not overcome the shortcomings of the classical theory of radiating charged particles. 

As we knew upfront that the classical theory is not able to treat correctly the radiation reaction force acting onto charged particles, which are radiating due to acceleration by electromagnetic forces, we could not reasonably expect that this problem would magically disappear when gravitation comes into play. While it is somewhat disturbing, that the gravitative radiation reaction force is not working onto the emitter but onto the receiver of the radiation, this is not worse than other oddities we are used to encounter with the classical model of radiating charges. Hence it would not be reasonable, now to shift the problem of radiation reaction to GRT, and doubt the validity of the EP. 

Instead clearly classical electrodynamics is responsible for this problem. We know that electromagnetic energy is emitted and absorbed not in form of continuous waves, but in form of photons. Hence quantum electrodynamics, but not classical electrodynamics, is appropriate to handle any questions regarding emission and absorption of radiation. If we want --- in spite the mentioned reservations --- to stick to the classical model, then we can summarize our conclusions, regarding a charge supported at rest in a gravitational field, as follows: 
\begin{ggitemize}
\ggitem{Observers, who are passing by in free fall (or otherwise constantly accelerated), see the charge radiating. When absorbing that radiation, they are decelerated. Thereby the gravitational field is loosing exactly the same amount of energy as is gained by absorption of radiation.}
\ggitem{Observers, who are at rest relative to the charge, don't see any radiation. They see that absorbers, which are passing by in free fall (or otherwise constantly accelerated), miraculously are absorbing electromagnetic energy out of nothing and at the same time are decelerating, such that the loss of gravitational energy is exactly balanced by the gain of electromagnetic energy. The strength of this effect is inversely proportional to the square of the distance between charge and absorber.}
\ggitem{It's not reasonable to speculate about radiation which is not absorbed by some absorbers, the more so as such radiation would violate energy conservation.}
\end{ggitemize} 
With the last sentence, we adopted the point of view of quantum theory, which is discussing observable phenomena, but not the point of view of classical physics, which is discussing alleged objective facts even if they are trivially not observable due to lack of an observer. 

\section{Free falling charges} 
As an immediate consequence of Larmor's formula\,\eqref{ksahghsdg} with the application note\,\eqref{eq:interpret}, a charged particle does not radiate if it is free falling in a gravitational field, because the gravitation is exerting the same acceleration on the particle and it's field, and hence there is no stress between the particle and it's field. We remark that inhomogeneities of the gravitational field (like tidal effects or anisotropy) will be negligible in the immediate vicinity of a charged particle (the near\bz{-}field region), and thus not induce radiation of free falling charges. 

\section{The consistency and beauty of theories} 
The theories of physics must comply with all observed phenomena (this is the requirement of correctness), they must provide an appropriate description of all observable phenomena (this is the requirement of completeness), and there must be no contradictions between various parts of physical theories (this is the requirement of consistency). 

The apparent conflict between GRT (respectively it's integral part, the EP) and classical electrodynamics, described in section\,\ref{sec:contradict}, was an inconsistency between theories. The conflict could be removed by a very slight adjustment of electrodynamics, \ggie  the clarification of Larmor's formula due to the application note\,\eqref{eq:interpret}, while GRT stayed untouched. That's not surprising. Remember for example that the conflict between GRT and quantum field theory, regarding the energy of the vacuum (the \al cosmological constant problem\arp ) could be settled\!\cite{Gruendler:ccp} by a slight modification of quantum field theory, while again GRT stayed untouched. If GRT is in conflict with some other theory, then most likely GRT will \al win\arp , while the other theory will need amendment. 

The exceptional strength of GRT is caused by the exceptional clarity and simplicity of it's premises. We could say: The exceptional strength of GRT is caused by it's beauty. In search for truth, the beauty of physical theories is a reliable guideline.\enlargethispage{2\baselineskip} 

{\interlinepenalty=100000{}\flushleft{\bibliography{../gg}}} 
\end{document}